\documentclass[12pt]{iopart}
\usepackage{color}
\usepackage{graphicx}
\usepackage[T1]{fontenc}
\usepackage{ae,aecompl}
\usepackage[english]{babel}
\usepackage{hyperref}
\usepackage{amssymb}

\newcommand{\ekin}{E_\mathrm{kin}}
\newcommand{\mdmin}{\rm md_{min}}
\newcommand{\net}{\mathrm{net}}
\newcommand{\diff}{\mathrm{diff}}
\newcommand{\lbra}{\left\langle}
\newcommand{\rket}{\right\rangle}
\newcommand{\vid}[1]{\lbra #1\rket}

\newlength{\figa}
\begin{document}


\title{Topological defect motifs in two-dimensional Coulomb clusters}
\author{A Radzvilavi\v{c}ius and E Anisimovas}
\address{Department of Theoretical Physics, Vilnius University,
Saul\.{e}tekio 9, LT-10222 Vilnius, Lithuania}
\date{\today}

\begin{abstract}
The most energetically favourable arrangement of low-density electrons 
in an infinite two-dimensional plane is the ordered triangular Wigner 
lattice. However, in most instances of 
contemporary interest one deals instead with finite clusters of strongly 
interacting particles localized in potential traps, for example, in
 complex plasmas. 
In the current contribution we study distribution of topological defects in two-dimensional 
Coulomb clusters with parabolic lateral confinement. The minima 
hopping algorithm based on molecular dynamics is used to 
efficiently locate the ground- and low-energy metastable states, 
and their structure is analyzed by means of the Delaunay triangulation. 
The size, structure and distribution of geometry-induced lattice 
imperfections strongly depends on the system size and the energetic 
state. Besides isolated disclinations and dislocations, classification
of defect motifs includes defect compounds --- grain boundaries, 
rosette defects, vacancies and interstitial particles. Proliferation 
of defects in metastable configurations destroys the orientational 
order of the Wigner lattice.
\end{abstract}


\maketitle

%
\section {Introduction}

The most energetically favourable arrangement of low-density electrons 
in an infinite two-dimensional plane is the ordered triangular Wigner 
lattice \cite{Wigner,Crandall} where each electron is surrounded by 
six equidistant nearest neighbours. However, in most instances of 
contemporary interest one deals instead with finite clusters of strongly 
interacting particles localized in potential traps. For example, 
extensively studied realizations of finite-size two-dimensional Wigner 
crystallites are created by trapping charged micrometric particles 
formed in the complex plasma environment \cite{Bonitz}. Confined 
geometries inevitably lead to deviations from the perfect triangular 
structure.

Presence of lattice imperfections plays a key role in determination of 
physical properties of both macroscopic \cite{Ashcroft} and low-dimensional 
mesoscopic systems. It has been argued that lattice faults are important 
in the shape changing processes during the maturation of capsides of 
spherical viruses \cite{Iorio}, cell division, and growth of bacterial 
surface layers \cite{Pum}. Intrinsic defects have a major impact on the 
collective particle motion and the inhomogeneous melting of two-dimensional 
Coulomb clusters in circular confinement \cite{Lai} and straight narrow 
channels \cite{Liu}.

The topological charge --- also known as the disclination charge --- of 
a site in a distorted lattice is defined as $Q = \tilde{C} - C$; here 
$C$ is the actual coordination number (the number of nearest neighbours) 
and $\tilde{C}$ is the reference value pertaining to the perfect lattice. 
For sites that belong to the bulk of the cluster $\tilde{C} = 6$, while 
for those situated on the edge $\tilde{C} = 4$. The topological charges 
of all sites in a two-dimensional cluster must sum up to six, a 
topological invariant due to the Euler's theorem \cite{Koulakov}. 

In a recent work \cite{Wales1} topological defects and defect complexes
were investigated in the Thomson model \cite{Thomson}, featuring a fixed 
number of particles moving on the surface of a sphere and interacting 
through repulsive Coulomb forces. In a quasi-three-dimensional geometry 
defined by the spherical surface the total disclination charge must equal 
 twelve \cite{Morris}. It was shown that in systems consisting of $400$ 
particles or more, isolated defects cost too much strain energy, and thus, 
various defect complexes are formed. The most intriguing of the lot are 
the so-called {\em rosette} defects consisting of a central 
five-coordinated site (topological charge $Q = 1$) surrounded by a layer 
of five seven-coordinated sites ($Q = -1$), and a further layer of five 
five-coordinated sites. The total disclination charge of this compound 
is $Q_\net = 1$. Thus, highly symmetric configurations with $12$ rosettes 
arranged on a sphere must be possible and were indeed reported in 
reference \cite{Wales1}.

The purpose of the present paper is a comprehensive study and
classification of topological defects in
a two-dimensional Coulomb cluster --- a mesoscopic system, known for its 
applicability to a broad range of experimental problems, namely, the 
distribution of electron islands on the surface of liquid helium 
\cite{Crandall,Rousseau}, electrons in semiconductor quantum dots 
\cite{Filinov} and charged micrometric spheres in the complex plasma 
environment \cite{Bonitz,Juan}. The system consists of a given number of 
identical charged particles confined by an isotropic parabolic potential 
trap. The confinement gives rise to the presence of topological 
defects. The number, structure and distribution of defects, or defect 
complexes, strongly depends on the system size and its energetic state. 

As a matter of fact, in large Coulomb clusters an exponentially large 
number of stable configurations are possible \cite{Radz}. The one with 
the lowest energy is called the ground state and is, of course, the most 
special and interesting configuration. However, metastable states are 
also realized in experiments \cite{Kahlert} and should not be neglected. 
Therefore, we do not restrict our attention to the ground states but 
investigate the whole spectrum, mostly focusing on its low-energy part.

We use the term \emph{low-energy states} to refer to the states whose
energies fall within the bottom $20\,\%$ of the energy interval covered
by all discovered metastable states.

The presence of a large number of competing metastable states is also a
computational issue. Various methods have been tried striving to locate the 
ground state as well as metastable configurations of large clusters. 
Simulated annealing \cite{NR}, genetic algorithms \cite{Morris}
and basin hopping \cite{Wales-BH} are a few significant examples. 
In the present work, we rely on the recently proposed minima hopping 
method \cite{Goedecker1}. This algorithm is designed to avoid revisiting 
already known configurations and employs low-energy molecular dynamics 
escape trajectories to reach neighbouring energy minima from a given one. 
The minima hopping method was shown to be efficient in locating 
ground states of complex molecular systems \cite{Goedecker2}, and with 
the present contribution we test its performance on strongly coupled 
Coulomb systems. 

The paper is organized as follows. In \Sref{model}, we introduce the 
model and the computational scheme. The three subsections of
\sref{motifs} are devoted to discussion and classification of the
obtained defect compounds and their impact on the orientational order
of the Wigner lattice. We end with a concluding \sref{concl}.

%
\section{Model system and computational approach}
\label{model}

The model system consists of $N$ identical particles of mass $m$ and 
electrostatic charge $q$, restricted to move in a two-dimensional plane 
and laterally confined by a circular parabolic potential. The total 
potential energy of the system is given by
\begin{equation}
\label{eq:potential}
U(\bi{r}_1, \ldots, \bi{r}_N)
= \sum_{i=1}^N \frac{1}{2} m\omega_0^2\bi{r}_i^2
+ \sum_{i>j}^N \frac{q^2}{r_{ij}}.
\end{equation}
Here, $\bi{r}_i$ is a two-dimensional vector denoting the position 
of the $i$-th particle and $r_{ij} = |\bi{r}_i-\bi{r}_j|$ is the 
distance between two of them. The first term of \eref{eq:potential} 
represents the parabolic confinement with the characteristic frequency 
$\omega_0$, while the second term corresponds to the Coulomb interparticle 
repulsion. Introducing the units of length $r_0=({q^2}/{m\omega_0^2})^{1/3}$ 
and energy $E_0={q^2}/{r_0}$, we rewrite the potential energy 
\eref{eq:potential} in a simple dimensionless and parameter-free form
\begin{equation}
\label{eq:dimlesspotential}
  U(\bi{r}_1, \ldots, \bi{r}_N)
  = \sum_{i=1}^N \frac{1}{2}\bi{r}_i^2
  + \sum_{i>j}^N \frac{1}{r_{ij}}.
\end{equation}
To find the low-energy stable states of the cluster, we use the minima 
hopping algorithm \cite{Goedecker1}. Since application of this algorithm 
is novel in this context, we begin with a brief outline.

Initially, the system is placed into one of the known minima and the 
algorithm starts by performing a short molecular dynamics simulation 
that attempts to escape from the basin of attraction of the current minimum. 
The particles are assigned random velocities according to the Maxwell 
distribution scaled in such a way that their total kinetic energy adds up 
to the current value of $\ekin$, a dynamically adjustable simulation
parameter. The molecular dynamics simulation proceeds 
keeping track of the instantaneous value of the potential energy, 
and is stopped immediately after its $\mdmin$-th minimum along the 
simulation path is reached. In the simplest version of the algorithm, the 
parameter $\mdmin$ is set to unity, that is, the nearest minimum is sought. 
However, one may often wish to skip some minima along the trajectory, and 
such an opportunity is provided by using larger values of $\mdmin$. 

Once the molecular dynamics simulation is stopped, the geometry 
minimization takes over and the system is forced to roll down towards 
the closest minimum. This step is accomplished by a combination of the
steepest descent and Newton's optimization techniques. 

There are three possible outcomes of the described simulation. If the 
kinetic energy $\ekin$ is insufficient to overcome the surrounding
potential barriers and/or the parameter $\mdmin$ is not large enough to 
find a low-energy escape path, the system rolls back to the same minimum. 
In the second case, the new minimum turns out to be visited previously. 
If any of these two outcomes is reached, the kinetic energy of the system 
is increased by a respective factor $\beta_1$ or $\beta_2$ (see below) 
and a new more vigorous attempt is launched from the same minimum. The 
third and most desirable case is realized when the obtained minimum is 
a new one and has not been visited previously. 

To introduce a preference for steps that reduce the total potential energy, 
the new configuration is accepted only if the increase in the potential 
energy does not exceed the threshold value $E_\diff$, another dynamically
adjustable parameter. If the newly found minimum is accepted, the kinetic 
energy of the system $\ekin$ is lowered and the following cycle is started 
from the new minimum.

The simulation is stopped after $\ekin$ increases significantly. This 
happens when the simulation is unable to produce any new minima for a
long time, which implies that the lowest minima have been found. 	

The behaviour of the energies $\ekin$ and $E_\diff$ is governed by five 
control parameters. The factors $\alpha_1$ and $\alpha_2$ determine how 
fast $E_\diff$ is increased (decreased) after a new minimum is rejected 
(accepted). The factors $\beta_1$, $\beta_2$ and $\beta_3$ multiply the 
kinetic energy $\ekin$ according to the outcome of an escape attempt, and 
must be chosen close to unity in order to sample the phase space 
thoroughly. In our simulations we follow \cite{Goedecker1} and use 
$\beta_1=\beta_2=1/\beta_3=1.05$ and $\alpha_1=1/\alpha_2=1.02$. 
Finally, $\mdmin$ determines the number of minima along the 
molecular dynamics path before the geometry relaxation takes over. 

The success of the algorithm is based on the validity of the 
Bell-Evans-Polanyi principle \cite{Goedecker1} which states that 
highly exothermic chemical reactions have a low activation energy. 
In other words, low-energy local minima are more likely to be located 
behind low potential barriers, and this is the reason to keep the kinetic 
energy as low as possible. Having performed some test runs we verified 
that the minima hopping algorithm never fails to find the true ground 
state of a three-dimensional Coulomb cluster with $N \leqslant 170$ 
particles and the Thomson problem with $N \leqslant 190$. 

We further investigated the performance of the minima hopping algorithm 
in the case of a two-dimensional $85$-particle cluster. This cluster has 
a symmetric ground state that controls a rather small basin of attraction
and $177$ metastable states known from the previous work \cite{Radz}.
Thus, the search for the ground state is not a simple task.

The algorithm starts at a randomly chosen initial minimum and runs until 
the true ground state is reached. To collect reliable statistics, the 
whole simulation is repeated $500$ times. The total number of the algorithm 
steps $n_{\rm steps}$ needed to reach the ground state strongly depends 
on the value of $\mdmin$ and correlates with the average kinetic energy 
of the system $\vid{\ekin}$ during the course of the simulation. From 
\tref{Table:MH} we observe that larger values of the parameter $\mdmin$ 
result in lower values of $\vid{\ekin}$ and a significant decrease of the 
total number of simulation steps. 

Higher values of $\mdmin$ are beneficial as they allow the system to 
oscillate more frequently within the area of the current basin or jump 
over a few barriers and thus increases chances to find low-energy 
escape trajectories. Higher values of $\mdmin$ also lead to a slightly 
lower number of minima $n_{\rm min}$ visited before the ground state is 
found. These observations confirm the validity of the Bell-Evans-Polanyi 
principle in Coulomb systems. On the other hand, large values of $\mdmin$
also increase the total computing time which strongly depends on the length 
of molecular dynamics trajectories. Therefore, in the bulk of our 
calculations we used moderate values $\mdmin = 2$ or $3$.

\Table{\label{Table:MH}The total number of minima hopping steps 
$n_{\rm steps}$, the number of visited minima $n_{\rm min}$ and the 
trajectory-averaged kinetic energy $\vid{\ekin}$ before the ground 
state is reached for different values of the control parameter $\mdmin$ 
in the 85-particle cluster.}
\begin{tabular}{@{}llll}
\br
$\mdmin$ & $n_{\rm min}$ & $n_{\rm steps}$ & $\vid\ekin$  \\
\mr
1 & 11.94 & 204.34 & 18.19 \\
10 & 12.05 & 135.30 & 7.51 \\
20 & 11.59 & 124.69 & 6.30 \\
25 & 10.27 & 104.20 & 5.06 \\
35 & 9.62 & 96.65 & 4.55 \\
50 & 9.44 & 93.14 & 4.03 \\
75 & 9.01 & 86.44 & 4.24 \\
90 & 8.58 & 80.86 & 3.80 \\
\br
\end{tabular}
\endTable 

%
\section{Defect motifs in two dimensions}
\label{motifs}

In this section, we report the results of the minima hopping simulations 
of two-dimensional classical Coulomb clusters in the range of sizes
$1 \leqslant N \leqslant 1000$ as well as a few cluster configurations 
of larger sizes: $N = 2\,000$, $3\,000$ and $4\,000$. The structure of 
the cluster and, consequently, the nature and distribution of topological 
defects strongly depends on the system size. Various aspects of
topological defects of relatively small clusters have been considered 
before \cite{Koulakov,Lai2,Kong}. Our aim is a comprehensive study
of a broad range of sizes, and most importantly, classification of
the determined structures some of which were previously not seen in 
two-dimensional systems.

\begin{figure}[ht]
\centering
\includegraphics[width=0.8\figa]{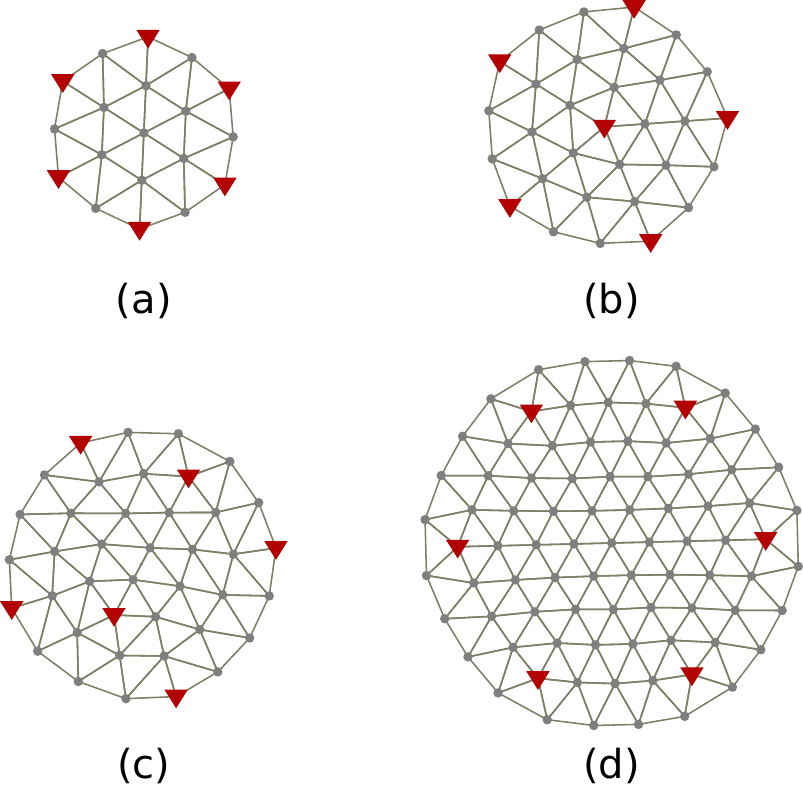}
\caption{Configurations of Coulomb clusters featuring six isolated 
disclinations denoted by red triangles: (a) the ground state of 
$19$-particle cluster, (b) a metastable state of a $31$-particle 
cluster, (c) a ground state of a $40$-particle cluster, and (d) the 
ground state of a $85$-particle cluster.}
\label{19-85}
\end{figure}

The analysis of the structure of a Coulomb cluster starts by determining 
the coordination number (the number of nearest neighbours) for all 
particles in the cluster. By definition, two particles are considered
nearest neighbours if they are connected by a link in the Delaunay
triangulation \cite{Geometry}, which provides an unambiguous algorithm 
to construct a net of triangles given a set of points on a plane. 
\Fref{19-85} provides several examples of triangulated cluster
structures. The symbols used in our figures are as follows: Gray lines 
depict the triangulation net, and ordinary six-coordinated vertices are 
shown by gray circles. Red triangles mark positive disclinations, 
that is, particles missing a neighbour; likewise blue squares (absent 
from \fref{19-85}) mark negative disclinations. Incidentally, the 
Delaunay triangulation is dual to the Voronoi construction 
\cite{Geometry} which is also often used in similar contexts
\cite{Lai2,Kong,Mughal}.

Distribution of defects in a parabolic confinement is by large determined
by a conflict between the circular boundary and the bulk-like interior
where a hexagonal lattice is preferred. Small clusters ($N \lesssim 70$) 
have no bulk and exhibit a shell structure defined by the circular symmetry 
of the confinement. A large fraction of the particles is located on the 
edge. Hence, small clusters often satisfy the Euler's theorem solely by 
the presence of edge disclinations, such as the $19$-particle cluster 
shown in \fref{19-85}~(a), while the distribution of defects is
sensitive to addition or subtraction of a single particle. 
Defect complexes are also frequently found 
in the interior, however, these defects are also the consequence of the 
well expressed shell structure and thus should not be interpreted as 
perturbations of a hexagonal lattice.

\begin{figure}[ht]
\centering
\includegraphics[width=0.9\figa]{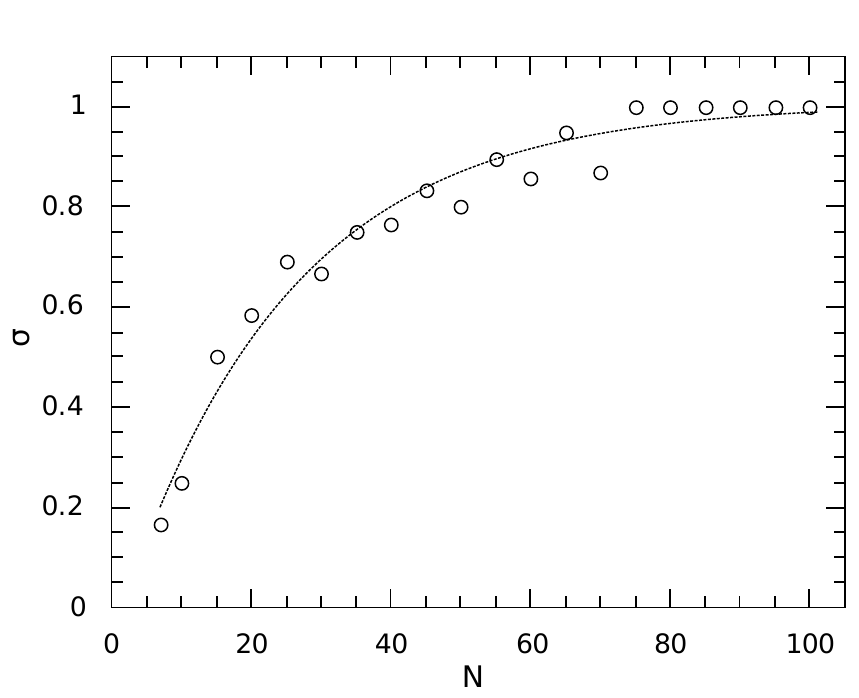}
\caption{Distribution of particles on the periphery of the ground-state 
clusters. The ratio of the number of particles in the outermost shell
to the number of particles in second farthest shell, $\sigma$, is 
plotted as a function of the system size $N$. The dotted line is drawn 
to guide the eye.}
\label{two-shells}
\end{figure}

\Fref{two-shells} shows that the number of particles in the two outermost 
shells equalizes for clusters larger than $N \approx 70$. Once this 
critical size is exceeded, a well-defined bulk develops, and only two or 
three external shells are nearly circular and contain approximately the 
same number of particles. In most cases, the defects move into the 
transition region between the circular outer part and the triangular 
interior. Six groups of defects, located at the corners of a hexagon can 
often be distinguished \cite{Kong}, and the size and composition of 
the observed defect structures become insensitive to the addition of a 
single particle to the cluster.

%
\subsection{Defect chains}

The predominant defect structure in two-dimensional Coulomb clusters 
is a defect chain consisting of alternating positive and negative 
disclinations, that is, five- and seven-coordinated vertices. In many 
(but not all) cases the number of positive disclinations in such a chain 
equals the number of negative disclinations plus one. Thus, the net 
topological charge of the chain equals $Q_\net = 1$.

A trivial case of this class is a chain of length one, that is, a single 
isolated defect. 

\textbf{Isolated positive disclinations} are frequently observed in 
relatively small systems. The presence of six $Q = 1$ defects is the 
simplest way to satisfy the Euler's theorem, however, stable 
configurations with six isolated disclinations, are not very common. 
A few notable high-symmetry examples are shown in \fref{19-85}.
Here we show two structures of
$C_6$ symmetry --- the ground states of $N = 19$ and 
$N = 85$ particle clusters in panels (a) and (d). Structures of $C_5$ 
symmetry are also possible, for example, the ground state of a 
$16$-particle cluster (not shown) and a metastable state of a $31$-particle 
cluster shown in panel (b). \Fref{19-85}~(c) shows an example of a 
low-symmetry configuration --- the ground state of a $40$-particle 
cluster.

\textbf{Isolated negative disclinations} are observed in some metastable 
configurations of small clusters, for example with $N=38$, $40$, $60$, 
and $80$. This defect, however, is always accompanied by a positive 
disclination or other defects in the close neighbourhood.

As the number of particles increases, isolated disclinations tend to 
disappear. Free disclinations are only rarely observed in low-energy 
states of clusters with the number of particles close to $N\approx300$ 
and are not observed at all in larger systems. 

\begin{figure}[ht]
\centering
\includegraphics[width=0.9\figa]{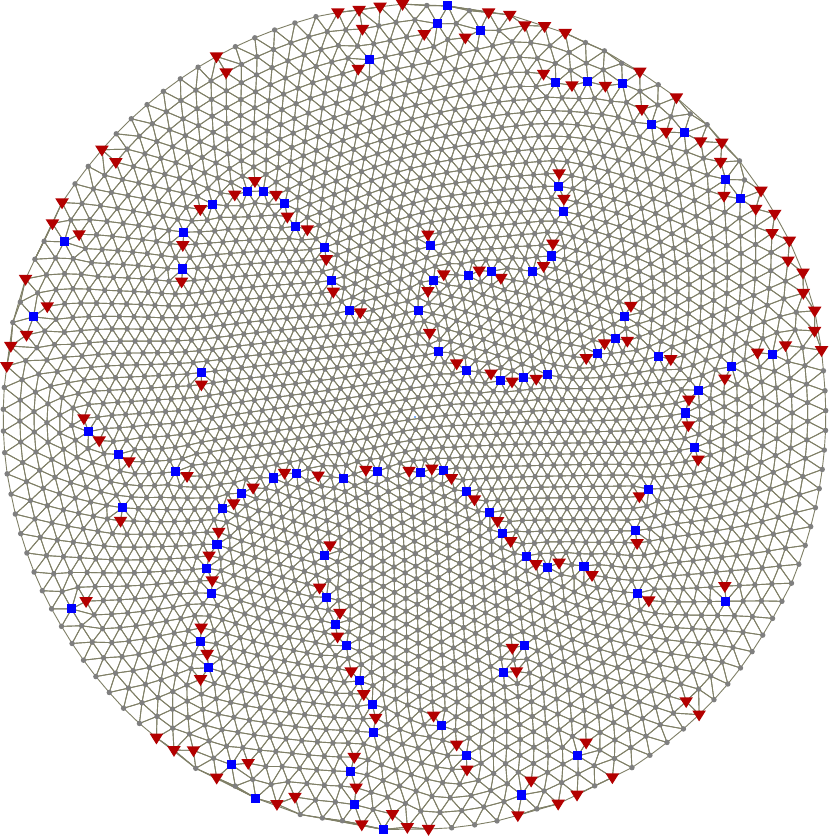}
\caption{Metastable state of a $3\,000$-particle Coulomb cluster. Long 
defect chains divide the interior into zones with different lattice 
orientation. Configurations of other investigated systems with several 
thousand particles are qualitatively similar.}
\label{GB}
\end{figure}

\textbf{A dislocation} is a topologically neutral defect, a pair of 
adjacent five- and seven-coordinated vertices, which may be regarded as 
a primitive chain of length two. Dislocations reduce strain induced by 
disclinations \cite{Koulakov}. The number of free dislocations in general 
increases with the energy of metastable state. Numerous dislocations can 
be seen in \fref{GB} which shows a high-energy metastable state of a 
$3\,000$-particle Coulomb cluster.

\begin{figure}[ht]
\centering
\includegraphics[width=0.9\figa]{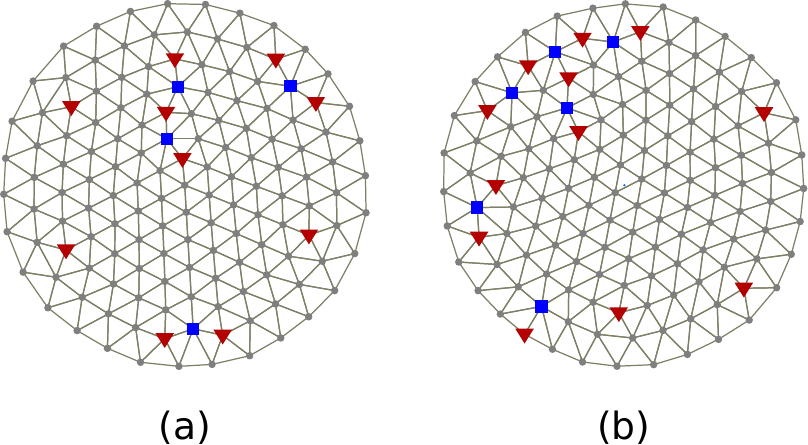}
\caption{Ground (a) and high-energy metastable (b) states of a
$143$-particle cluster. In the ground state, three disclinations, two 
extended dislocations and short grain boundary occupy the six corners 
of a hexagon. The hexagonal structure is lost in metastable states.}
\label{143}
\end{figure}

In larger systems, longer chains of defective vertices become energetically 
favourable. \textbf{An extended dislocation}, also referred to as a scar
\cite{Wales1}, is a negative disclination flanked by two positive
disclinations. It has the shape close to that of the $\rm{H_2O}$ molecule 
[see \fref{143}~(a)] and the net topological charge $Q_\net = 1$. 
This structure may be regarded
as a rudimentary grain boundary. Scars are often present in metastable 
states of small clusters and ground states in the size range of 
$N \approx 100\,\ldots\, 460$ particles. Configurations with six isolated 
extended dislocations at the six corners of hexagon are observed at some 
values of $N$; a few examples are $N=200$, $210$, $250$, and $N=291$. In 
larger clusters, such as the near-ground-state of $N=560$ system, scars 
appear in complexes with additional dislocations.

\textbf{Length-5} grain boundaries, consisting of 
alternating three $Q = 1$ and two
$Q = -1$ vertices appear in metastable states of small clusters. Later, 
these defect chains replace single disclinations or extended dislocations 
in some of the six defect groups of large clusters. An illustrative 
example with all three types of defects is given in \fref{143}. 
As the number of particle grows, the number and length of the grain 
boundaries in the low-energy metastable states on the average increases.
However, the growth of the number of defective vertices is not uniform.
In many cases of large clusters, at least one of these defect chains 
points radially towards the center of the cluster; the others separate 
the inner region from the circular part of the cluster (see \fref{520}). 
Some of the grain boundaries are also accompanied by a few five- and 
seven-coordinated vertex pairs.

\textbf{Lengthy grain boundaries} and neutral disclination-dislocation 
chains appear in the interior of high-energy metastable states of large 
systems. They divide the inner part of the cluster into distinct zones 
with different orientations of crystallographic axes. Extremely long 
defect chains ($20$ vertices and more) are present in metastable 
states of crystals with a few thousand particles shown in \fref{GB}, 
giving rise to polycrystalline order.

%
\subsection{Complex structures}

A number of defect structures do not follow the general shape of a chain.
The first of these complex motifs is a \textbf{twin grain boundary} with 
a local mirror line. The twin grain boundary consists of two negative 
disclinations, sharing a common triangulation edge, and three positive
disclinations on the periphery. Thus, the net topological charge of 
the structure is $Q_\net = 1$. Twin grain boundaries have been reported 
in the low-energy minima of the Thomson model \cite{Wales2} but were not 
discussed in 2D Coulomb systems. 

\begin{figure}[ht]
\centering
\includegraphics[width=0.6\figa]{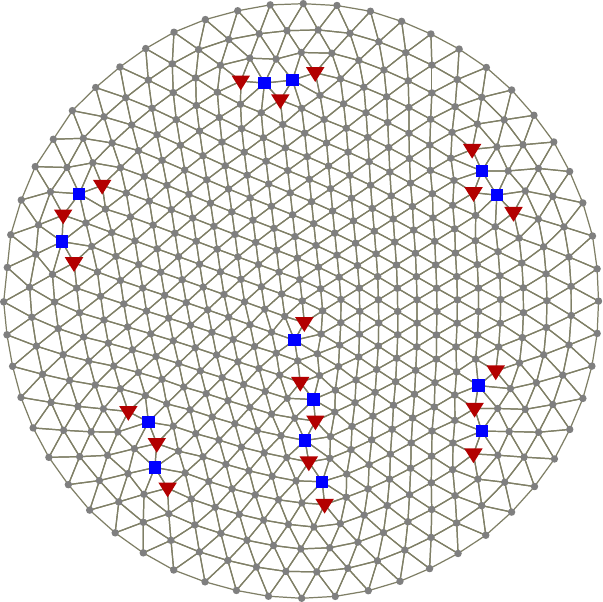}
\caption{Low-energy metastable state of $520$-particle two-dimensional 
Coulomb cluster shows a typical defect distribution of this size range. 
Two twin grain boundaries, three length-5 grain boundaries and one radial 
length-7 defect chain, screened by an additional dislocation are located 
in the corners of a hexagon.}
\label{520}
\end{figure}

\begin{figure}[ht]
\centering
\includegraphics[width=0.9\figa]{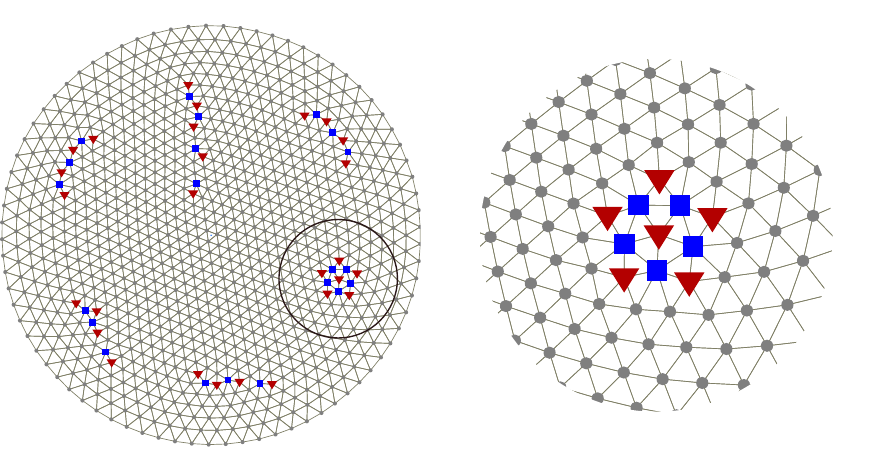}
\caption{Rosette defect in low-energy metastable state of a 
$1\,000$-particle cluster.}
\label{N1000-1}
\end{figure}

Twin grain boundaries first appear in metastable states of relatively small 
clusters ($N=90$ is one of the examples), and then among the groups of 
defects situated in the corners of a hexagon in larger systems as seen in
\fref{520}. Some twin grain boundaries in large crystals are extended by 
a few additional pairs of $Q = \pm1$ defects, or are screened by 
neighbouring dislocations (see \fref{N1000-1}).

A unique \textbf{rosette} defect of pentagonal symmetry is present in 
some of the low-energy metastable states of large clusters ($N>600$). A 
rosette consists of a central positive disclination surrounded by five 
negative disclinations alternating with five positive disclinations.
The net topological charge of this complex equals $Q_\net = 1$. Highly 
symmetric configurations of $12$ rosettes were previously observed in 
low-energy states of spherical Thomson clusters \cite{Wales1}. In 
two-dimensional Coulomb systems rosettes were not reported before, and 
we observe at most one rosette per cluster. Further simulations of 
larger crystals would be needed to investigate, if the presence of 
multiple rosettes is ever possible.

\begin{figure}[ht]
\centering
\includegraphics[width=\figa]{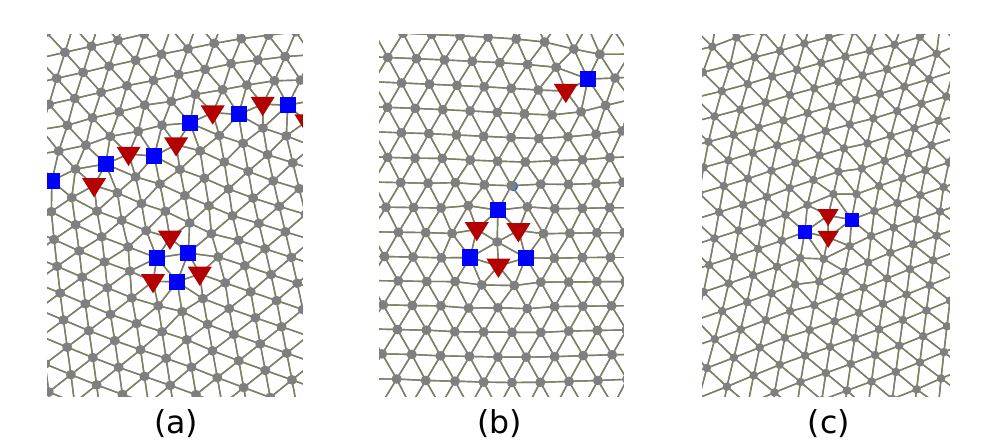}
\caption{Topologically neutral complexes: (a) a triangle-shaped vacancy 
in the core of high energy metastable state of $N=900$ cluster situated 
close to a long grain boundary, (b) an interstitial defect in the 
inner region of $N=690$ Coulomb system, and (c) a rhombus-shaped 
interstitial in a metastable state of a $N=604$ particle cluster.}
\label{Triangles}
\end{figure}

A few classes of cyclic defect configurations are observed in high-energy 
metastable states of large systems and are shown in \fref{Triangles}.
There are two triangular structures composed of three positive and three
negative disclinations, and a rhombus-shaped structure consisting of two 
$Q = 1$ vertices accompanied with two $Q = -1$ vertices. Two types of 
triangular structures are present: Panel (a) shows the smaller of the two,  
with three seven-coordinated particles moved close to the center of the 
motif, and panel (b) presents the larger structure which encircles a
six-coordinated particle. All these complexes are topologically neutral 
and do not change the global hexagonal symmetry of the lattice. However, 
they do change the local density of the particles --- the complex shown 
in \fref{Triangles}~(a) is a vacancy, whereas the ones seen in
\fref{Triangles}~(b) and (c) are interstitial particles \cite{Candido}.

%
\subsection{Bond-orientational order}

As the energy of a metastable state increases, the average number of 
defective vertices grows \cite{Kong} and the hexagonal symmetry of the 
defect clustering positions is lost. Topological defects are observed 
both in the interior and on the edge of a cluster. Lattice imperfections 
tend to form long chains and asymmetric complexes that heavily distort 
the triangular structure. To quantify the loss of structural order in 
high-energy metastable states, we evaluate the site-averaged complex 
bond-orientational order (BOO) parameter. For a single site, the BOO 
parameter is defined as \cite{Halperin} 
\begin{equation}
\Psi_6   = \frac{1}{C} \sum_{k=1}^{C} 
 \exp({\rmi 6 \theta_{k}}).
\end{equation}
Here, $\theta_k$ is the angle of the vector drawn from the particle in
question to the $k$-th of its nearest neighbours. An ordered site on a 
perfect triangular lattice has $|\Psi_6|=1$. A slightly distorted site 
with six nearest neighbours and low strain usually has $|\Psi_6|>0.4$,
whereas a heavily distorted site with five or seven neighbours has 
$|\Psi_6|<0.4$ \cite{Liu}. To evaluate the degree of orientational 
order in the core of a cluster, we average the absolute value of the 
BOO parameter $\Psi_6$ over all sites except those located within three 
outermost shells.

The lowest-energy configurations of large clusters correspond to the 
bond-orientational order parameter within the range 
$|\Psi_6| \approx 0.66\,\ldots\, 0.80$, depending on the actual size and 
distribution of defects with respect to the center of the cluster. 
For example, a low-energy state of the $520$-particle crystal shown in
\fref{520} includes a radially oriented compound of a length-7 grain 
boundary and a nearby dislocation, and thus has a low value of 
$|\Psi_6|=0.66$. On the other hand, the longest defect chain in the 
system with $N=550$ particles is composed of only five defective 
vertices. The core of the cluster is defect free, and hence, the 
orientational order is relatively high, $|\Psi_6|=0.8$. 

\begin{figure}[ht]
\centering
\includegraphics[width=\figa]{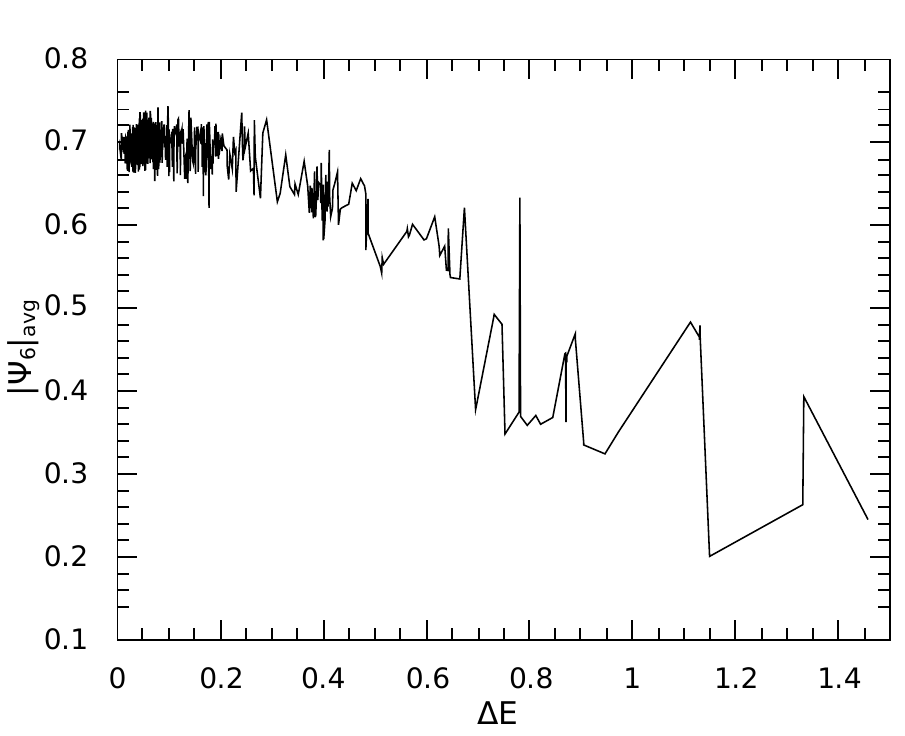}
\caption{Absolute value of bond-orientational order parameter $\Psi_6$ 
averaged over the sites at the core of the 2D Coulomb cluster with 
$N=1000$ particles versus the relative energy of metastable states. 
Here, $\Delta E$ is the energy measured from the lowest found state.}
\label{Psi}
\end{figure}

The dependence of the averaged BOO parameter $|\Psi_6|_{\rm{avg}}$ on 
the energy of the metastable state is shown in \fref{Psi} for $N=1\,000$. 
We see, that at low energies the values of the BOO parameter indicate 
ordered states of the lattice ($\Psi_6 \approx 0.7$). The value of the 
order parameter is subject to substantial fluctuations, however, the general
trend is clearly visible: As the energy of the state increases, the 
order is gradually lost. The values of BOO drop as low as 
$|\Psi_6|_{\rm{avg}} \approx 0.2$ for metastable states with the highest 
energies. These configurations have a large number of long grain 
boundaries at the core of the cluster that separate regions of different
local orientation of the lattice.

We observe that metastable states of large clusters, different in size 
or orientation of only a few defects, are very close in their energy. 
As a notable example we consider a metastable state of $N=1\,000$ 
particle system and energy $E=103\,457.1505$. A slight change of the 
orientation of a defect pair near the extended dislocation changes the 
energy of the state by $\Delta E = 0.0002$. Likewise, as the disclination 
screened by two dislocations transforms into length-5 grain boundary, 
the energy of the system increases by only $0.0001$.

%
\section{Conclusion}
\label{concl}

We performed minima hopping simulations of isotropic
two-dimensional Coulomb clusters in order to study the distribution 
and transformations of intrinsic topological defects in low-energy
stable configurations. We conclude, that with properly chosen values 
of control parameters, the minima hopping technique is an efficient 
and fast method to locate the candidates for the ground states in 
strongly coupled Coulomb systems.  

The distribution and size of the geometry-induced defects strongly 
depends on the system size and its energetic state. As the cluster 
size increases, free disclinations are gradually replaced by more 
energetically favourable extended dislocations and grain boundaries 
of non-uniformly increasing length. A novel rosette defect, previously 
seen only in metastable states of a quasi-three-dimensional Thomson 
model, is shown to be present in low-energy states of large clusters. 
However, we observe only one rosette per crystal. Cyclic defects --- 
triangle and rhombus-shaped vacancies and interstitial particles --- 
are observed in high-energy metastable states of two-dimensional 
Coulomb systems.

As the energy of metastable states grows, the orientational order at 
the core of large clusters is lost due to the appearance of elongated 
grain boundaries and increased concentration of dislocations. 

\ack
This research was funded by a grant No.\ MIP-79/2010 from the Research 
Council of Lithuania.

\end{document}